\documentstyle[12pt]{article}
\begin{document}
\newcommand{\cl}{\centerline}
\renewcommand{\theequation}{\arabic{equation}}
\newcommand{\beq}{\begin{equation}}
\newcommand{\eeq}{\end{equation}}
\newcommand{\bea}{\begin{eqnarray}}
\newcommand{\eea}{\end{eqnarray}}
\newcommand{\nn}{\nonumber}
\newcommand\pa{\partial}
\newcommand\un{\underline}
\newcommand\ti{\tilde}
\newcommand\pr{\prime}
\begin{titlepage}
\setlength{\textwidth}{5.0in} \setlength{\textheight}{7.5in}
\setlength{\parskip}{0.0in} \setlength{\baselineskip}{18.2pt}

\begin{flushright}
{\tt hep-th/0401170} \\
{\tt SOGANG-HEP 306/03}
\end{flushright}

\vspace*{0.3cm}

\begin{center}
{\large\bf Symplectic Embedding of a Massive Vector-Tensor Theory
with Topological Coupling}
\end{center}

\begin{center}
{Yong-Wan Kim$^{}$\footnote{\small Electronic address:
ywkim@sejong.ac.kr} and Chang-Yeong Lee$^{}$\footnote{\small
Electronic address: cylee@sejong.ac.kr} \\
Department of Physics and Institute for Science and
Technology,\\
Sejong University, Seoul 143-747, Korea\\
Seung-Kook Kim$^{}$\footnote{\small Electronic address:
skandjh@empal.com}\\
{Department of Physics, Seonam University, Chonbuk 590-170, Korea\\
Young-Jai Park$^{}$\footnote{\small Electronic address:
yjpark@sogang.ac.kr}}\\
Department of Physics, Sogang University, Seoul 121-742, Korea}\\
\end{center}

\begin{center}
(\today)
\end{center}
\vfill

\vskip 0.4cm

\begin{center}
{\bf ABSTRACT}
\end{center}
\begin{quotation}
In the symplectic Lagrangian framework we newly embed an
irreducible massive vector-tensor theory into a gauge invariant
system, which has become reducible, by extending the configuration
space to include an additional pair of scalar and vector fields,
which give the desired Wess-Zumino action. A comparision with the
BFT Hamiltonian embedding approach is also done.

\vskip 0.4cm

\end{quotation}
\end{titlepage}

\newpage
\section{Introduction}

The Dirac quantization method \cite{dirac64} has been widely used
in order to quantize Hamiltonian system involving first- and
second-class constraints. However, the resulting Dirac brackets
may be field-dependent and nonlocal, and thus pose serious
ordering problems. The BRST quantization \cite{becci76} on the
lines of Dirac established by Batalin, Fradkin, and Vilkovisky
\cite{fradkin75}, and then improved in a more tractable and
elegant Hamiltonian embedding by Batalin, Fradkin, and Tytin (BFT)
\cite{batalin87}, does not suffer from these difficulties, as it
relies on a simple Poisson bracket structure. As a result, this
embedding of second-class system into first-class one has received
much attention in the past decade \cite{nwz,phyrep,KKP}.

While the various quantization methods based on the Hamiltonian
formulation are developed for general types of gauge theories,
Faddeev and Jackiw \cite{jackiw85} have introduced an equivalent
scheme based on a first order Lagrangian which does not need to
classify constraints as primary, secondary, etc. Since this scheme
deals with less number of constraints than that of Dirac, it is
proved to be relatively simple to find the Dirac brackets. After
this work, there are numerous related analyses
\cite{wozneto,symplectic,sym-string}. In particular, we had shown
that in this framework, a gauge non-invariant theory can be
embedded into a gauge invariant one by investigating the
properties of the symplectic two-form matrix and its corresponding
zero modes \cite{kpkk}. Recently, there are some renewed interests
\cite{sym-emb,hkp,sym-emb-new} in the subjects of symplectic
embedding including noncommutative theories.

On the other hand, antisymmetric tensor fields, which first appear
as a mediator of the string interaction \cite{kalb}, have been
much interested in an alternative of the Higgs mechanism
\cite{djt,alt-higgs}. With the topologically interacting term of
the form $B\wedge F$, the physical degree of freedom of the
antisymmetric rank two tensor field $B$ is absorbed by the vector
field, making it massive. This mechanism is considered generic in
string phenomenology \cite{string}. Moreover, various dual
descriptions between different theories have been widely studied
where the antisymmetric tensor field plays an important role in
realization of dualities \cite{dual,banerjee}. These dual
relations were also independently confirmed by defining dual
operations on the space of pairs of different gauge forms
\cite{cantcheff}. These were a surprising result that, as like the
well-known dual equivalence \cite{dj} of the first order self-dual
theory and Maxwell-Chern-Simons theory in $d=3$, they have all
shown that the first order master action even in $d=4$ has a dual
relation with the Maxwell-Kalb-Ramond (MKR) theory. Recently, it
has shown that through the BFT embedding technique the gauge
non-invariant master action is equivalent to the gauge invariant
MKR theory \cite{harikumar}. Very recently, we have generalized
their results to include both the gauge symmetry breaking term and
the topological coupling one \cite{KKP}, resulting in a new type
of Wess-Zumino (WZ) action \cite{faddeev86} as well as the usual
St\"uckelberg one.

This paper is organized as follows. In section 2, we briefly show
the dual relation of the first order master Lagrangian between the
Abelian Proca and the Kalb-Ramond (KR) massive theories
classically by using the equations of motion. Then, we explicitly
carry out the symplectic quantization for this gauge noninvariant
Lagrangian. In section 3, we embed this master Lagrangian into a
desired gauge invariant one, and make clear the relation between
the ``trivial" zero modes and the symmetry transformations. We
also explicitly show that one of the ``trivial" zero modes is
related to a reducible constraint. In section 4, we compare the
result of the symplectic embedding with the BFT one. Our
conclusion is given in section 5.

\section{Symplectic Quantization of the Master Lagrangian}
\setcounter{equation}{0}
\renewcommand{\theequation}{\arabic{section}.\arabic{equation}}

In this section, we reconsider the first order Lagrangian
\cite{banerjee} for the Abelian massive vector and tensor fields,
and quantize it explicitly through the symplectic scheme. This
Lagrangian is not only known to have the dual relation with the
MKR Lagrangian \cite{KKP,banerjee,cantcheff,harikumar}, but also
plays a role of a master Lagrangian of the Proca and the KR
models.

First, let us briefly review that the following first order
Lagrangian
\begin{eqnarray}
\label{master-action}
 {\cal L} =-\frac{1}{4} B_{\mu\nu} B^{\mu\nu}
             + \frac{1}{2}  A_\mu A^{\mu} + \frac{1}{2 m}
             \epsilon_{\mu \nu \rho \sigma}
             B^{\mu \nu} \partial^\rho A^\sigma
\end{eqnarray}
describes classically the massive Proca and KR theories
simultaneously, {\it i.e.,} the Lagrangian (\ref{master-action})
is a master Lagrangian\footnote{For a quantum mechanical
treatment, see the Ref.\cite{banerjee}.} of the mentioned two
theories. By varying the Lagrangian with the fields $A^\mu$ and
$B^{\mu\nu}$, one obtains their equations of motion as
\begin{eqnarray}
\label{eqAB}
 B_{\mu \nu} - \frac{1}{m} \epsilon_{\mu \nu \rho \sigma}
 \partial^\rho A^\sigma  &=& 0,
\nonumber \\
A_\mu + \frac{1}{2 m} \epsilon_{\mu \nu \rho \sigma} \partial^\nu
B^{\rho \sigma} &=& 0,
\end{eqnarray}
respectively. By eliminating the antisymmetric $B^{\mu\nu}$ fields
from the Lagrangian (\ref{master-action}), the Proca model is
obtained as follows
\begin{eqnarray}
\label{proca} {\cal L}_{Proca} = - \frac{1}{4 m^2} F_{\mu \nu}
F^{\mu \nu} + \frac{1}{2} A_\mu A^\mu.
\end{eqnarray}
Similarly, by using the equation of motion for the field $A^\mu$,
one could obtain the KR Lagrangian as
\begin{eqnarray}
\label{KR} {\cal L}_{KR} = \frac{1}{12 m^2} G_{\mu \nu \sigma}
G^{\mu \nu \sigma} - \frac{1}{4} B_{\mu \nu} B^{\mu \nu}.
\end{eqnarray}
This shows that the first order Lagrangian (\ref{master-action})
is the master Lagrangian of the two massive theories at the
classical level.

On the other hand, according to the symplectic scheme
\cite{jackiw85,wozneto}, it is easy to find their symplectic
brackets in a few steps equivalent to the Dirac ones in the
Hamiltonian formulation as
\begin{eqnarray}
\label{Proca_pb}
\{ A^0 (x) , A^i (y) \} &=& \partial_x^i
\delta(x-y) ,\nonumber \\
\{ A^i (x), \pi_j (y) \} &=& m^2 \delta_j^i \delta(x-y)
\end{eqnarray}
for the Proca model \cite{hkp} as well as
\begin{eqnarray}
\label{K-R pb}
\{ B^{0i} (x), B^{jk} (y) \} &=& ( \delta^{ij}
\partial_x^k - \delta^{ik} \partial_x^j ) \delta (x-y)
\nonumber , \\
\{ B^{ij} (x) , \pi_{kl} (y) \} &=& m^2 ( \delta_k^i \delta_l^j -
\delta_k^j \delta_l^i ) \delta (x-y)
\end{eqnarray}
for the KR theory \cite{BC}.

Now, in order to implement the usual symplectic quantization of
the master Lagrangian itself which is composed of two kinds of the
fields with different ranks as well as their topological coupling
term, we rewrite the Lagrangian  to an alternating first-ordered
one from symmetrized form of Eq. (\ref{master-action}) as
\begin{equation}
\label{1st-lag} {\cal L}^{(0)} = \frac {1}{4 m} \epsilon_{ijk}
B^{jk}{\dot A}^i - \frac{1}{4 m} \epsilon_{ijk} A^k {\dot B}^{ij}-
{\cal H}^{(0)},
\end{equation}
where the Hamiltonian is
\begin{eqnarray}
\label{1st-ham} {\cal H}^{(0)} = \frac{1}{4} B_{\mu\nu} B^{\mu\nu}
- \frac{1}{2} A_\mu A^\mu - \frac{1}{m} \epsilon_{ijk} B^{0i}
\partial^j A^k - \frac {1}{2 m} A^0 \epsilon_{ijk}
\partial^i B^{jk}.
\end{eqnarray}
Here, we adopt the conventions: $\epsilon_{0ijk}=\epsilon_{ijk}$,
$\epsilon^{123}=+1$, and $g^{\mu\nu}=(+---)$. As is clear in the
Lagrangian (\ref{1st-lag}), there are no needs to introduce
additional auxiliary fields such as momenta because it already has
the form of first order.

Then, we identify the initial sets of symplectic variables
and their conjugate momenta from the first order Lagrangian as
follows
\begin{eqnarray}
\label{ini-vars}
\xi^{(0)\alpha} &=& ( A^i, B^{ij}, A^0, B^{0i}), \nonumber \\
a^{(0)}_\alpha &=& (\frac{1}{4m}\epsilon_{ijk}B^{jk},
-\frac{1}{2m}\epsilon_{ijk}A^k, 0, \vec{0}^T).
\end{eqnarray}
Note that the coefficients of the fields having time derivative in
the canonical sector play the roles of momenta, and $B^{ij}$ in
the symplectic variable $\xi^{(0)\alpha}$ denote the independent
component fields such as $B^{ij}:=(B^{12}, B^{23}, B^{31})$ in
order, collectively, while any contracted indices are understood
to be summed over unless otherwise mentioned. From the definition
of the symplectic two-form matrix \cite{wozneto}:
\begin{equation}
\label{symp-tensor} f_{\alpha\beta}(x,y) = \frac {\partial a_\beta
(y)}{\partial \xi^\alpha (x)} - \frac {\partial a_\alpha
(x)}{\partial \xi^\beta (y)},
\end{equation}
we explicitly obtain the following zeroth-iterated matrix
\begin{equation}
\label{ini-symp-matrix} f_{\alpha\beta}^{(0)} (x,y)  = \left(
\begin{array}{cccc}
O & \frac{1}{m}\epsilon_{i(jk)} & \vec {0} & O\\
-\frac{1}{m}\epsilon_{(ij)k} & O & \vec {0} & O\\
\vec {0}^T & \vec {0}^T &  0 & \vec {0}^T\\
O & O & \vec {0} & O \\
\end{array}\right)\delta(x-y).
\end{equation}
The tensor components $\epsilon_{i(jk)}, \epsilon_{(ij)k}$ are
given by
\begin{eqnarray}\label{epsilon-0ij}
\epsilon_{i(jk)}= \left(
\begin{array}{ccc}
  0 & 1  & 0 \\
  0 & 0 & 1 \\
  1 & 0 & 0\\
\end{array} \right),
\end{eqnarray}
where the $i$/$jk$-components outside/inside the parenthesis
denote explicitly the vector/tensor fields, and behave as like the
totally antisymmetric tensor $\epsilon_{ijk}$, similar for
$\epsilon_{(ij)k}$. Thereinafter, let us omit the parenthesis
otherwise a confusion arises. We also denote that in the matrix
$f_{\alpha\beta}^{(0)}$ the $O$, $\vec{0}$ and $\vec{0}^T$ symbols
stand for a $3\times 3$ null matrix, a three-dimensional column
null vector, and its transpose, respectively. Since we easily know
that the matrix $f_{\alpha\beta}^{(0)}$ is singular, there exist
four-fold infinities related to zero modes,
$\tilde{\nu}^{(0)T}_{\alpha}(\sigma,x)$, labelled by discrete
indices $\sigma=(\epsilon_1,\vec{\epsilon}_2)$ with
$\vec{\epsilon}_2\equiv \left((\epsilon^1_2
,0,0),(0,\epsilon^2_2,0),(0,0,\epsilon^3_2)\right)$, where
$\epsilon_1$ and $\vec{\epsilon}_2$ are arbitrary functions of the
continuum label $x$, explicitly with components
\begin{eqnarray}
\label{zeromode1} \tilde{\nu}^{(0)T}_{\alpha}(\epsilon_1,x)&=&
(\vec{0}, \vec{0}, \epsilon_1, \vec{0}), \nonumber \\
\tilde{\nu}^{(0)T}_{\alpha}(\vec{\epsilon}_2,x)&=& (\vec{0},
\vec{0}, 0, \vec{\epsilon}_2).
\end{eqnarray}
Therefore, the four-fold zero modes
$\tilde{\nu}^{(0)T}_{\alpha}(\sigma,x)$ generate four
corresponding Lagrangian constraints $\Omega_{\epsilon_1}$ and
$\Omega_{\vec{\epsilon}_2}$ defined as
\begin{eqnarray} \int dx ~\Omega_\sigma (x) = \int dx
~\tilde{\nu}^{(0)T}_{\alpha}(\sigma,x)
\frac{\partial}{\partial\xi^{(0)\alpha}(x)}\int dy~{\cal
H}^{(0)}(y)=0, \nonumber
\end{eqnarray}
such that
\begin{eqnarray}
\label{lag-cons}  \Omega_{\epsilon_1} &=& A_0 + \frac{1}{2 m}
\epsilon_{ijk} \partial^i B^{jk}=0, \nonumber\\
\Omega_{\vec{\epsilon}_2} &=& B_{0i} - \frac{1}{m} \epsilon_{ijk}
\partial^j A^k=0.
\end{eqnarray}
These constraints should be conserved in time, which requirement
is incorporated into the Lagrangian (\ref{1st-lag}), resulting in
an extension of the symplectic space with auxiliary fields,
$\alpha$, $\beta^i$, which correspond to Lagrange multipliers.

As a result, the first iterated Lagrangian is written as
\begin{equation}
\label {iterated-lag} {\cal L}^{(1)} = \frac {1}{4m}
\epsilon_{ijk} \dot{A}^i B^{jk} - \frac{1}{4m} \epsilon_{ijk}
\dot{B}^{ij} A^k + \Omega_{\epsilon_1} \dot{\alpha} +
\Omega_{\epsilon^i_2} \dot{\beta}^i -{\cal H}^{(1)},
\end{equation}
where the first-iterated Hamiltonian is given by
\begin{eqnarray}
\label {iterated-ham} {\cal H}^{(1)} (\xi) &=& {\cal H}^{(0)}
|_{\Omega_\sigma = 0} \nonumber \\
&=& \frac{1}{4} B_{ij} B^{ij} - \frac{1}{2} A_i A^i - \frac{1}{2}
B_{0i} B^{0i} + \frac{1}{2} A_0 A^0.
\end{eqnarray}

We have now for new symplectic variables and their conjugate
momenta as
\begin{eqnarray}
\label{1st-iterated-var} \xi^{(1)\alpha} &=& \left \{A^i, B^{ij},
A^0, B^{0i}, \alpha, \beta^i \right \},
\nonumber\\
a^{(1)}_\alpha &=& \left \{ \frac{1}{4m} \epsilon_{ijk} B^{jk} ,
-\frac{1}{2m}\epsilon_{ijk} A^k, 0, \vec{0}^T,
\Omega_{\epsilon_1}, \Omega_{\epsilon^i_2} \right \}.
\end{eqnarray}
The first-iterated symplectic matrix is again obtained as
\begin{equation}\label{1st-iterated-symp-matrix}
f_{\alpha\beta}^{(1)} (x,y)  = \left(
\begin{array}{cccccc}
O & \frac{1}{m}\epsilon_{ijk} & \vec {0} & O & \vec {0} &
\frac{1}{m}\epsilon_{ijk}\partial^k\\
-\frac{1}{m}\epsilon_{ijk} & O & \vec {0} & O & -\frac{1}{m}
\epsilon_{ijk}\partial^k & O\\
\vec {0}^T & \vec {0}^T &  0 & \vec {0}^T & 1 & \vec {0}^T\\
O & O & \vec {0} & O & \vec {0} & -\delta_{ij}\\
\vec {0}^T & -\frac{1}{m}\epsilon_{ijk}\partial^k & -1 &
\vec {0}^T & 0 & \vec {0}^T\\
\frac{1}{m}\epsilon_{ijk}\partial^k & O & \vec {0} & \delta_{ij}
& \vec {0} & O\\
\end{array}\right)\delta(x-y).
\end{equation}

One can now easily see that this matrix has an inverse as
\begin{equation}\label{inverse-symp-matrix}
\left( f_{\alpha\beta}^{(1)} \right)^{-1} (x,y)  = \left(
\begin{array}{cccccc}
O & -m\epsilon_{ijk} & \partial^i & O & \vec{0} & O\\
m\epsilon_{ijk} & O & \vec{0} & F_{24} & \vec{0} & O\\
\partial^i & \vec{0}^T &  0 & \vec{0}^T & -1 & \vec{0}^T\\
O & F_{24}^T & \vec{0} & O & \vec{0} & \delta_{ij}\\
\vec{0}^T & \vec{0}^T & 1 & \vec{0}^T & 0 & \vec{0}^T\\
O & O & \vec{0} & -\delta_{ij} & \vec{0} & O\\
\end{array}\right)\delta(x-y),
\end{equation}
where
\begin{equation}
\label{f24} F_{24} = \left (
\begin{array}{ccc}
\partial^2 & - \partial^1 & 0\\
0 & \partial^3 & -\partial^2 \\
-\partial^3 & 0 & \partial^1\\
\end{array} \right).
\end{equation}
Since there are no more new non-trivial zero modes, the iteration
process is stopped at this stage. From this inverse matrix we
easily read the desired nonvanishing symplectic brackets for the
vector and the tensor fields as
\begin{eqnarray} \label{brackets}
\left \{ A^0 (x), A^i (y) \right \} &=& \partial_x^i \delta( x- y)
, \nonumber \\
\left \{ A^i (x), B^{jk} (y) \right \} &=& m \epsilon^{ijk}
\delta( x- y) , \nonumber \\
\left \{ B^{0i} (x), B^{jk} (y) \right \} &=& \left( \delta^{ij}
\partial_x^i  - \delta^{ik} \partial_x^j \right ) \delta( x- y).
\end{eqnarray}
This ends the usual symplectic procedure with the only four true
constraints in Eq. (\ref{lag-cons}).

Comparing this symplectic scheme with the Dirac formulation, one
easily sees that the former has less number of the constraints
than that of the latter which has 14 constraints shown in Eqs.
(\ref{pri-con}) and (\ref{sec-con}) in section 4, does not need to
define primary constraints, and thus more efficient to get the
brackets. In this respect, it is generally understood that the
symplectic method deals only with true constraints while Dirac's
one over-constrained.

\section{Gauge Invariant Symplectic Embedding}
\setcounter{equation}{0}
\renewcommand{\theequation}{\arabic{section}.\arabic{equation}}

In this section, we embed the first order master Lagrangian
without resorting to the Hamiltonian method, which is the usual
way of getting corresponding gauge invariant Lagrangian. In the
usual BFT embedding scheme, one first works with full Dirac
constraints and Hamiltonian defined in the phase space, converts
them into gauge invariant ones through the systematic BFT scheme
in an extended phase space, and finally using the path integral
methods performs a series of integrations for momentum variables
to find out a corresponding gauge invariant Lagrangian in an
extended configuration space. Compared with this rather long
procedure, the gauge invariant symplectic embedding scheme has its
merit of simplicity based on the singular property of the
symplectic matrix with corresponding ``trivial" zero modes which
will be defined shortly.

The idea is simply to consider that a desired gauge invariant
Lagrangian resulting from an embedding procedure would be provided
by
\begin{equation}
\label{lag-tot} {\cal L}_{T}={\cal L}_O + {\cal L}_{WZ},
\end{equation}
where the Lagrangian ${\cal L}_O$ is the symmetrized original
master Lagrangian ${\cal L}^{(0)}$ in Eq. (\ref{1st-lag}). Then,
the symplectic procedure is greatly simplified, if we make the
following ``educated" guess for WZ Lagrangian\footnote{Similar to
the Proca model \cite{hkp}, one could consider the Ansatz more
general and manifestly Lorentz invariant functions of WZ
Lagrangian composed of a scalar $\theta$ and vector fields
$Q^\mu$, however those consideration only adds unnecessary
calculational complication. On the other hand, in recent works
\cite{sym-emb-new}, there makes an attempt on the Lagrangian
embedding a bit systematic, however in practice it seems very
complicated even in the simplest one-form cases
\cite{kpkk,sym-emb,hkp} including the Proca model. Furthermore, it
may be intractable to apply the method to reducible systems,
including the massive vector-tensor theory with topological
coupling.}, respecting the Lorentz symmetry,
\begin{equation}
{\cal L}_{WZ}= c_1 \partial_\mu\theta\partial^\mu\theta
         +c_2 A^{\mu}\partial_\mu \theta + c_3 Q_{\mu \nu}
         Q^{\mu \nu} + c_4 B_{\mu \nu} Q^{\mu \nu},
\label{lagwz}
\end{equation}
where $Q_{\mu \nu}= \partial_\mu Q_{\nu} -
\partial_\nu Q_{\mu}$. Here $\theta$ and $Q^\mu$ are gauge degrees
of freedom with respect to the original fields $A^\mu$ and
$B^{\mu\nu}$, respectively. As an Ansatz for a consistent Lorentz
covariant embedding, we shall take the coefficients $c_1, c_2,
c_3, c_4$ to be constants, and we will fix these by the condition
that a finally iterated symplectic matrix has ``trivial" zero
modes on the one hand and these zero modes should generate no new
constraints on the other hand. Furthermore, we will show that the
``trivial" zero modes related with the finally iterated symplectic
matrix correctly yield proper rules of gauge transformation
including a first-stage reducible trivial zero mode associated
with a reducible constraint.

Now, we could read the canonical momenta of the total Lagrangian
as
\begin{eqnarray}
\label{momenta} && \pi_0 = -c_2\theta,~~~~~~~ \pi_{\theta}
=2c_1\dot\theta,
\nonumber\\
&& \pi_{0i}=-2c_4 Q_i,~~~P_i = 4c_3 (\dot{Q}_i-\partial_i Q_0).
\end{eqnarray}
Here, we have used the partially integrated Lagrangian for the
second and fourth terms in Eq. (\ref{lagwz}) in order to easily
show the coincidence with the constraints obtained from the BFT
embedding in section 4. Along with these auxiliary variables
(momenta), to implement the symplectic procedure as like in the
previous section, we write the total Lagrangian (\ref{lag-tot}) in
the first-ordered form as
\begin{equation}
\label{1st-lag-emb} {\cal L}_{T}^{(0)} = \frac{1}{4m}
\epsilon_{ijk} B^{jk}\dot{A}^i  -
\frac{1}{4m}\epsilon_{ijk}A^k\dot{B}^{ij} + \pi_0\dot{A}^0 +
\pi_{\theta}\dot\theta + \pi_{0i}\dot{B}^{0i}+P_i {\dot Q}^i -
{\cal H}_{T}^{(0)},
\end{equation}
where the Hamiltonian is given by
\begin{eqnarray}
\label{1st-hamil} {\cal H}_{T}^{(0)} &=& \frac{1}{4} B_{\mu\nu}
B^{\mu\nu} - \frac{1}{2} A_\mu A^\mu - \frac{1}{m} \epsilon_{ijk}
B^{0i} \partial^j A^k - \frac {1}{2m} A^0 \epsilon_{ijk}
\partial^i B^{jk}
\nonumber\\
&+&\frac{1}{4c_1}\pi_\theta^2 +\frac{1}{8c_3}P_i P^i
-c_1\partial_i\theta\partial^i\theta-c_2A^i\partial_i \theta
-c_3Q_{ij}Q^{ij}-c_4B_{ij}Q^{ij}\nonumber \\
&+& P_i \partial^i Q^0+2c_4B^{0i}\partial_iQ_0.
\end{eqnarray}

Then, the initial set of symplectic variables $\xi^{(0)\alpha}$
and their conjugate momenta $a^{(0)}_\alpha$ are given by
\begin{eqnarray}
\xi^{(0)\alpha}&=&(A^i,B^{ij},\theta,\pi_\theta,Q^i,P_i,A^0,B^{0i},Q^0),\nonumber\\
a^{(0)}_\alpha&=&(\frac{1}{4m}\epsilon_{ijk}B^{jk},-\frac{1}{2m}\epsilon_{ijk}A^k,
\pi_\theta,0,P_i,\vec{0}^T,-c_2\theta,-2c_4 Q_i,0). \label{a0wz}
\end{eqnarray}
From the above set of the symplectic variables we read off the
symplectic matrix defined in Eq. (\ref{symp-tensor}) to be
\begin{equation}
f^{(0)}_{\alpha\beta}(x,y)= \left(
\begin{array}{ccccccccc}
O & \frac{1}{m}\epsilon_{ijk}&\vec{0} &\vec{0} & O &O&\vec{0} &O&\vec{0} \\
-\frac{1}{m}\epsilon_{ijk}&O&\vec{0}&\vec{0}&O&O&\vec{0}&O&\vec{0} \\
\vec{0}^{T} &\vec{0}^{T} &0 &-1 &\vec{0}^{T}&\vec{0}^{T}&-c_2&\vec{0}^{T}&0\\
\vec{0}^{T} &\vec{0}^{T} &1 &0  &\vec{0}^{T}&\vec{0}^{T}&0&\vec{0}^{T}&0\\
O&O&\vec{0}&\vec{0}&O&-\delta_{ij}&\vec{0}&2c_4\delta_{ij}&\vec{0}\\
O&O&\vec{0}&\vec{0}&\delta_{ij}&O&\vec{0}&O&\vec{0}\\
\vec{0}^{T} &\vec{0}^{T} &c_2 &0  &\vec{0}^{T}&\vec{0}^{T}&0&\vec{0}^{T}&0\\
O&O&\vec{0}&\vec{0}&-2c_4\delta_{ij}&O&\vec{0}&O&0\\
O&O&\vec{0}&\vec{0}&O&O&\vec{0}&O&\vec{0}\\
\end{array}
\right)\delta(x-y), \label{f0wz0}
\end{equation}
which is manifestly singular as we observe the last null row and
column, and thus the matrix has non-trivial zero modes given by
\begin{eqnarray}
\nu^{(0)T}_{\alpha}(\epsilon_1,x)&=&(\vec{0}^T, \vec{0}^T, 0,
-c_2 \epsilon_1,\vec{0}^T,\vec{0}^T, \epsilon_1,\vec{0}^T,0),\nonumber\\
\nu^{(0)T}_{\alpha}(\vec{\epsilon}_2,x)&=&(\vec{0}^T, \vec{0}^T,
0, 0,\vec{0}^T,2c_4 \vec{\epsilon_2}, 0,\vec{\epsilon_2},0),\nonumber\\
\nu^{(0)T}_{\alpha}(\epsilon_3,x)&=&(\vec{0}^T, \vec{0}^T, 0,
0,\vec{0}^T,\vec{0}^T, 0,\vec{0}^T,\epsilon_3).
\end{eqnarray}
Applying these zero-modes from the left to the equation of motion
we have obtained constraints $\phi_\sigma\equiv
(\phi_{\epsilon_1}, \phi_{\vec{\epsilon}_2}, \phi_{\epsilon_3})$
as
\begin{eqnarray}
\phi_{\epsilon_1}&=& A_0 +\frac{1}{2m}\epsilon_{ijk}
\partial^iB^{jk}+\frac{c_2}{2c_1}\pi_{\theta}=0, \nonumber\\
\phi_{\vec{\epsilon}_2}&=& B_{0i}-\frac{1}{m}\epsilon_{ijk}
\partial^jA^k-\frac{c_4}{2c_3}P_i=0, \nonumber\\
\phi_{\epsilon_3}&=& \partial^iP_i+2c_4\partial_i B^{0i}=0.
\label{constsym}
\end{eqnarray}

Next, following the symplectic algorithm for theories having gauge
symmetry \cite{wozneto,kpkk,hkp}, we obtain the first-iterated
Lagrangian by enlarging the canonical sector with the constraints
$\phi_\sigma$ and their associated Lagrangian multipliers
$(\alpha, \beta^i,\gamma)$, respectively, as follows
\begin{eqnarray}
\label{1st-iter-L} {\cal L}^{(1)}_{T}&=&\frac{1}{4m} \epsilon_{ijk}
B^{jk}\dot{A}^i - \frac{1}{4m}\epsilon_{ijk}A^k\dot{B}^{ij}
-2c_2\theta\dot{A}^0 + \pi_{\theta}\dot\theta
-2c_4Q_i\dot{B}^{0i}+ P_i {\dot Q}^i
\nonumber \\
&+&\phi_{\epsilon_1}\dot{\alpha}+\phi_{\vec{\epsilon}_2}\dot{\beta}^i
+\phi_{\epsilon_3}\dot{\gamma} - {\cal H}_{T}^{(1)},
\end{eqnarray}
where ${\cal H}_T^{(1)}={\cal H}^{(0)}|_{\phi_{\epsilon_3}=0}$ is
the first-iterated Hamiltonian. We have now for the first-level
symplectic variables $\xi^{(1)\alpha}$ and their conjugate momenta
$a^{(1)}_\alpha$
\begin{eqnarray}
\xi^{(1)\alpha}&=&(A^i,B^{ij},\theta,\pi_\theta,Q^i,P_i,A^0,B^{0i},\alpha,\beta^i,\gamma),
\nonumber\\
a^{(1)}_\alpha&=&(\frac{1}{4m}\epsilon_{ijk}B^{jk},-\frac{1}{2m}\epsilon_{ijk}A^k,
\pi_\theta,0,P_i,\vec{0}^T,0,\vec{0}^T,\phi_{\epsilon_1},
\phi_{\vec{\epsilon}_2},\phi_{\epsilon_3}),
\end{eqnarray}
and the first-iterated symplectic matrix now reads as
\begin{equation}
\label{emb-2form-mat} f^{(1)}_{\alpha\beta}(x,y) = \left(
\begin{array}{cc}
        f^{(0)}_{\hat{\alpha}\hat{\beta}} & M_{\hat\alpha\sigma}\\
        -M^{T}_{\sigma\hat\alpha} &O
\end{array}
\right)\delta(x-y),
\end{equation}
where the submatrix $f^{(0)}_{\hat\alpha\hat\beta}$ refers to the
$\xi_{\hat\alpha}=(A^i,B^{ij},\theta,\pi_\theta,Q^i,P_i,A^0,B^{0i})$
sector, and $M_{\hat{\alpha}\sigma}$ is a $18\times 5$ matrix
defined as $M_{\hat{\alpha}\sigma}=\frac{\partial\phi_\sigma(y)}
{\partial \xi_{\hat{\alpha}}(x)}$:
\begin{equation}
M_{\hat\alpha\sigma}(x,y)=\left(
\begin{array}{ccc}
\vec{0} &-\frac{1}{m}\epsilon_{ijk}\partial^k_y &\vec{0}\\
\frac{1}{m}\epsilon_{ijk}\partial^k_y &O &\vec{0}\\
0 & \vec{0}^{T} &0\\
\frac{c_2}{2c_1}&\vec{0}^T&0\\
\vec{0}&O&\vec{0}\\
\vec{0}&-\frac{c_4}{2c_3}\delta_{ij}&\partial^i_y\\
1&\vec{0}^T&0\\
\vec{0}&-\delta_{ij}&2c_4\partial^y_i\\
\end{array}
\right)\delta(x-y).
\end{equation}
Then, this symplectic matrix, which contains no null rows and
columns, is seemingly non-singular. Note that we observe the $Q^0$
term does not appear in the canonical sector of the Lagrangian
${\cal L}^{(1)}$ as well as in the Hamiltonian ${\cal H}^{(1)}_T$
due to the use of the constraint $\phi_{\epsilon_3}$. It is also
important to eliminate the $Q^0$ field in the first-iterated level
in the technical point of view. Keeping $Q^0$ field in the
Lagrangian ${\cal L}^{(1)}$ makes trouble in the symplectic matrix
because there is always zero row or column due to the absence of
its conjugate momentum in $a^{(1)}(\xi)$. It prevents us from
finding desired zero modes.

Now, the essential point of the symplectic Lagrangian embedding
\cite{kpkk,sym-emb,hkp,sym-emb-new} is this: in order to realize a
gauge symmetry in this approach, the matrix (\ref{emb-2form-mat})
must have at least one ``trivial" zero mode which does not
generate a new constraint. In our case, the solution of
\begin{eqnarray}
\int {\rm d}y~ f^{(1)}_{\alpha\beta}(x,y)\nu^{(1)}_\beta(y) = 0
\nonumber
\end{eqnarray}
yields the following ``trivial" zero modes
\begin{eqnarray}
&&\nu^{(1)T}_{\alpha}({\epsilon_1},x)=(\vec{0}^T, \vec{0}^T, 0,
-c_2{\epsilon_1}, \vec{0}^T, \vec{0}^T, {\epsilon_1}, \vec{0}^T,
0, \vec{0}^T, 0),
\nonumber\\
&&\nu^{(1)T}_{\alpha}(\vec{\epsilon}_2,x)=(\vec{0}^T, \vec{0}^T,
0, 0, \vec{0}^T, 2c_4\vec{\epsilon}_2, 0, \vec{\epsilon}_2, 0,
\vec{0}^T,0),\nonumber\\
&&\nu^{(1)T}_{\alpha}({\epsilon_3},x)=(\partial^i\epsilon_3,
\vec{0}^T, -\frac{1}{c_2}\epsilon_3, 0, \vec{0}^T, \vec{0}^T, 0,
\vec{0}^T, \epsilon_3,\vec{0}^T, 0), \nonumber\\
&&\nu^{(1)T}_{\alpha}(\epsilon^i_4,x)=(\vec{0}^T,
F^T_{24}\vec{\epsilon}_4, 0, 0, -\frac{1}{2c_4}\vec{\epsilon}_4,
\vec{0}^T, 0, \vec{0}^T, 0, \vec{\epsilon}_4, 0),\nonumber\\
&&\nu^{(1)T}_{\alpha}(\epsilon_5,x)=(\vec{0}^T, \vec{0}^T, 0, 0,
\partial_i \epsilon_5, \vec{0}^T, 0, \vec{0}^T, 0, \vec{0}^T,
\epsilon_5), \label{zero1}
\end{eqnarray}
where $F^T_{24}$ is defined in Eq. (\ref{f24}), while giving the
relations for the free adjustable coefficients
\begin{equation}
c^2_2=2c_1,~~~c^2_4=-c_3,\label{c1234}
\end{equation}
which are obtained from the single-valuedness of the zero mode
functions.

For consistency, we can confirm that these zero modes do not
really generate any new constraint provided we apply them to the
right-hand side of the equations of motion
\begin{eqnarray}
\int {\rm d}x~\nu^{(1)}_\alpha(\sigma,x)\frac{\partial}{\partial
\xi^{(1)\alpha}(x)}\int {\rm d}y~{\cal H}^{(1)}_T(y)
=0,~~~\sigma=(\epsilon_1,\vec{\epsilon}_2, \epsilon_3,
\vec{\epsilon}_4, \epsilon_5), \nonumber
\end{eqnarray}
explicitly as
\begin{eqnarray}
\label{1st-iter-zero-modes} \int {\rm
d}x~\nu^{(1)T}_{\alpha}(\epsilon_1,x) \frac{\delta}
{\delta\xi^{(1)\alpha}(x)} \int {\rm d}y~{\cal H}^{(1)}(y) &=&
-\int {\rm d}x~\epsilon_1\phi_{\epsilon_1}=0,
\nonumber\\
\int {\rm d}x~\nu^{(1)T}_{\alpha}(\vec{\epsilon}_2,x)
\frac{\delta}{\delta\xi^{(1)\alpha}(x)} \int {\rm d}y~{\cal
H}^{(1)}(y)&=& \int
{\rm d}x~\vec{\epsilon}_2\phi_{\vec{\epsilon}_2}=0, \nonumber\\
\int {\rm d}x~\nu^{(1)T}_{\alpha}(\epsilon_3,x)
\frac{\delta}{\delta\xi^{(1)\alpha}(x)} \int {\rm d}y~{\cal
H}^{(1)}(y)&=& \int
{\rm d}x~\epsilon_3(1-\frac{c_2^2}{2c_1})\partial_i A^i=0, \nonumber\\
\int {\rm d}x~\nu^{(1)T}_{\alpha}(\epsilon_4,x) \frac{\delta}
{\delta\xi^{(1)\alpha}(x)} \int {\rm d}y~{\cal H}^{(1)}(y) &=&-
\int
{\rm d}x~\epsilon^i_4(1+\frac{c_4^2}{c_3})\partial^j B_{ij}=0, \nonumber\\
\int {\rm d}x~\nu^{(1)T}_{\alpha}(\epsilon_5,x) \frac{\delta}
{\delta\xi^{(1)\alpha}(x)} \int {\rm d}y~{\cal H}^{(1)}(y)
&=&0, \nonumber\\
\end{eqnarray}
where the third and fourth equations are reconfirmed by using the
relations in Eq. (\ref{c1234}). Since the last equation concerning
the zero mode $\epsilon_5$ identically vanishes, this false zero
mode plays no role at all, reflecting that the symplectic scheme
deals only with true constraints.

Therefore, before proceeding further to the symplectic embedding
algorithm, with these determination of the coefficients we should
go back to the first-iterated level (\ref{1st-iter-L}) because the
constraints are now not all independent, {\it i.e.,} we can
identify the last two constraints in Eq. (\ref{constsym}) as
\begin{equation}
\partial^i \phi_{\epsilon^i_2} - \frac{1}{2c_4}\phi_{\epsilon_3}=0,
\end{equation}
which means a gauge invariant Lagrangian for what we are seeking
is reducible. In this case, in order for resolving the reducible
constraint, we have to modify the auxiliary field as $\beta^i
\rightarrow \beta^i-2c_4\partial^i \gamma$. Then, a pair of the
symplectic variable and the momentum, $(\gamma,
\phi_{\epsilon_3})$, are absorbed into the modified $\beta^i$
variables. This also modifies the symplectic matrix
(\ref{emb-2form-mat}) to a new one which does not have the last
row and column like the one in (\ref{f0wz0}). Finding solutions
having new zero modes is exactly equivalent in the corresponding
``trivial" zero modes (\ref{1st-iter-zero-modes}) to set the
$\epsilon_5$-parameter zero. This clearly explains why the last
zero mode in Eq. (\ref{zero1}) does not generate any new
constraints but vanishes identically.

On the other hand, we could also interpret these results in view
of linear dependence of the ``trivial" zero modes as follows: In
the zero mode solutions (\ref{zero1}) we have explicitly inserted
the gauge parameters $\epsilon_\sigma$ keeping the form of gauge
transformations (\ref{trfmrule}) in mind. Instead, since we could
normalize them by introducing the delta functional\footnote{We can
also label the zero modes as follows:
$\nu^{(l)}_{\alpha,y}(\sigma, x)$, $(\sigma=1,...,N)$, where
``$l$" refers to the ``level", $\alpha$, $y$ stand for the
component, while $\sigma$, $x$ label the $N$-fold infinity of zero
modes in ${\cal R}^3$. See more details in Ref. \cite{hkp}.} in
the symplectic embedding scheme \cite{hkp}, we then have the last
two zero modes in Eq. (\ref{zero1}) explicitly as
\begin{eqnarray}
\label{reducible-zero} &&\nu^{(1)T}_{\hat\alpha,
y}(\vec{\epsilon}_4,x)=(\vec{0}^T, F^T_{24},
0,0,-\frac{1}{2c_4}\vec{e},\vec{0}^T,0,\vec{0}^T)
\delta(x-y),\nonumber\\
&&\nu^{(1)T}_{\hat\alpha, y}(\epsilon_5,x)=(\vec{0}^T, \vec{0}^T,
0, 0,\nabla\epsilon_5, \vec{0}^T,0,\vec{0}^T)\delta(x-y),
\end{eqnarray}
where $\hat\alpha$ denote the component fields as
$\xi_{\hat\alpha}=(A^i,B^{ij},\theta,\pi_\theta,Q^i,P_i,A^0,B^{0i})$
and $\vec{e}$ is a unit 3-dimensional vector. Then, we easily see
that the zero mode concerning with the gauge parameter
$\epsilon_5$ is related to $\nu^{(1)T}_{\hat\alpha,
y}(\vec{\epsilon}_4,x)$ as
\begin{equation}
\nu^{(1)T}_{\hat\alpha, y}(\epsilon_5,x)=
-2c_4\nabla\nu^{(1)T}_{\hat\alpha, y}(\vec{\epsilon}_4,x)
\end{equation}
showing that it is not linearly independent, {\it i.e.,}
reducible. In this point of view, we would call it,
$\nu^{(1)T}_{\hat\alpha, y}(\epsilon_5,x)$, as a `first-stage
reducible' trivial zero mode. This extends previously known
results \cite{kpkk,sym-emb,hkp,sym-emb-new} in which the existence
of ``trivial" zero modes, or equivalently no new constraints are
generated in the symplectic framework, implies gauge symmetry in
system to include the first-stage reducible case.

As a result, we arrive at the final result on the symplectic
embedding of the massive vector-tensor theory with
topological coupling. The desired gauge invariant Lagrangian is
now explicitly written as
\begin{eqnarray}
\label{T-Lag}
{\cal L}_T &=& {\cal L}_O + {\cal L}_{WZ} \nonumber\\
&=& -\frac{1}{4}(B_{\mu\nu}-2c_4 Q_{\mu\nu})^2
+\frac{1}{2}(A_\mu+c_2\partial_\mu\theta)^2
+\frac{1}{2m}\epsilon_{\mu\nu\rho\sigma}B^{\mu\nu}
\partial^\rho A^\sigma.\nonumber\\
\end{eqnarray}
This ends the Lagrangian embedding in the symplectic setup.

Now, in order to discuss gauge transformation, we consider the
symplectic matrix (\ref{emb-2form-mat}). As has been shown in Ref.
\cite{wozneto,kpkk}, the ``trivial'' zero modes generate gauge
transformations on the symplectic variable
$\xi^{(1)}_{\hat\alpha}$ in the sense
\begin{equation}
\delta\xi_{\alpha}(x)=\Sigma_{\tilde\sigma}
\nu^{(1)T}_{\alpha}(\tilde\sigma,x),~~~\tilde\sigma=(\epsilon_1,\vec{\epsilon}_2,
\epsilon_3, \vec{\epsilon}_4). \label{trfmrule}
\end{equation}
We thus obtain the gauge transformations of the symplectic
variables from Eq. (\ref{zero1}) as follows
\begin{eqnarray}
\delta A^0 &=& \epsilon_1,~~~ \delta A^i =
\partial^i \epsilon_3,~~~
\delta \theta = -\frac{1}{c_2}\epsilon_3,\nonumber\\
\delta B^{0i} &=& \epsilon^i_2,~~~ \delta B^{ij}=
-(\partial^i\epsilon^j_4-\partial^j\epsilon^i_4),~~~ \delta
Q^i=-\frac{1}{2c_4}\epsilon^i_4,
\end{eqnarray}
except the missing transformation $\delta Q^0$ which cannot be
obtained at this stage due to the elimination of the $Q^0$
component in the first-iterated level of the procedure. This is
very similar with the BFT Hamiltonian embedding for the
constrained reducible system in which we should introduce the
$Q^0$ field in the path integral measure in order to keep manifest
covariance in the Lagrangian\footnote{We will discuss this
explicitly in the next section.}.

To find out explicitly the gauge transformation for the $Q^0$
field as well as any existence of possible restrictions to the
gauge parameters, let us consider the total variation of the
Lagrangian (\ref{T-Lag}), which transforms as
\begin{eqnarray}
\delta{\cal L}_{T} &=& (A_0+c_2\partial_0\theta)(\epsilon_1-
\partial^0\epsilon_3) \nonumber \\
&+& (\frac{1}{m}\epsilon_{ijk}\partial^j A^k-B_{0i}+2c_4Q_{0i})
(\epsilon^i_2+\partial^0\epsilon^i_4+2c_4\partial^i\delta Q^0).
\end{eqnarray}
Therefore, if we identify the gauge parameters as
\begin{eqnarray}
\epsilon_1 =\partial^{0}{\epsilon}_3,~~~
\epsilon^i_2=-(\partial^{0}{\epsilon}^i_4-\partial^i\epsilon^0_4),
\end{eqnarray}
along with a new parameter $\epsilon^0_4$ defined as
\begin{equation}
\delta Q^0 = -\frac{1}{2c_4}\epsilon^0_4,
\end{equation}
we can make the total Lagrangian ${\cal L}_{T}$ gauge invariant. As a result, this
Lagrangian is invariant under the final gauge transformations as
\begin{eqnarray}
\label{zeromode2} && \delta A^\mu = \partial^\mu\epsilon_3,
~~~\delta\theta =
-\frac{1}{c_2}\epsilon_3, \nonumber\\
&& \delta B^{\mu\nu} =
-(\partial^\mu\epsilon^\nu_4-\partial^\nu\epsilon^\mu_4),~~~
\delta Q^\mu = -\frac{1}{2c_4}\epsilon^\mu_4.
\end{eqnarray}
Therefore, by considering purely the symplectic Lagrangian
embedding, we have found the gauge invariant Lagrangian as well as
their full gauge transformations including $\delta Q^0$. Note that
the coefficients $c_2$, $c_4$ can be either rescaled on the
variables as $(\theta, Q^\mu)\rightarrow(c_2\theta, 2c_4 Q^\mu)$,
or fixed as $c_2=\pm 1, c_4=\pm\frac{1}{2}$. Then, the resulting
Lagrangian is reduced to the well-known gauge embedded form of the
massive vector-tensor theory with topological coupling,
\begin{eqnarray}
\label{T-Lag1} {\cal L}_T = -\frac{1}{4}(B_{\mu\nu}-
Q_{\mu\nu})^2+\frac{1}{2}(A_\mu+\partial_\mu\theta)^2
+\frac{1}{2m}\epsilon_{\mu\nu\rho\sigma}B^{\mu\nu}
\partial^\rho A^\sigma,
\end{eqnarray}
where we set the free adjustable coefficients as $c_2=1$ and
$c_4=1/2$, and this Lagrangian is invariant under
\begin{eqnarray}
&& \delta A^\mu = \partial^\mu\epsilon, ~~~\delta\theta =
-\epsilon, \nonumber\\
&& \delta B^{\mu\nu} =
\partial^\mu\epsilon^\nu-\partial^\nu\epsilon^\mu,~~~ \delta
Q^\mu = \epsilon^\mu,
\end{eqnarray}
where the gauge parameters are redefined as
$\epsilon=\epsilon_3$ and $\epsilon^\mu=-\epsilon^\mu_4$.

\section{Revisited BFT Embedding}
\setcounter{equation}{0}
\renewcommand{\theequation}{\arabic{section}.\arabic{equation}}

In this section, we will compare the symplectic Lagrangian
embedding with the previous work \cite{KKP} of the BFT Hamiltonian
one, which makes second-class constraint Hamiltonian system into
corresponding first-class one in a systematic way.

First, let us start with the canonical momenta from the
symmetrized form of the Lagrangian (\ref{master-action}) as $\pi_0
= 0$, $\pi_i = \frac{1}{4m}\epsilon_{ijk}B^{jk}$, $\pi_{0i} = 0$,
and $\pi_{ij} = - \frac{1}{2m}\epsilon_{ijk}A^k$ in order to
analyze the Hamiltonian structure of the model. Then, we have
obtained the primary Hamiltonian in the Dirac's terminology as
\begin{eqnarray}
\label{pri-H} {\cal H}_p &=&
\frac{1}{4}B_{\mu\nu}B^{\mu\nu}-\frac{1}{2}A_\mu A^\mu
-\frac{1}{m}\epsilon^{ijk}B_{0i}\partial_j
A_k-\frac{1}{2m}A^0\epsilon^{ijk}\partial_iB_{jk}\nonumber\\
&+& \lambda^0 \pi_0 + \lambda^i \Omega_i
+\Sigma^{0i}\pi_{0i}+\Sigma^{ij}\Omega_{ij},
\end{eqnarray}
where the ten primary constraints are given by
\begin{eqnarray}
\label{pri-con} && \pi_0\approx 0,
~~~\Omega_i\equiv\pi_i-\frac{1}{4m}\epsilon_{ijk}B^{jk}\approx 0,\nonumber\\
&& \pi_{0i}\approx 0,~~~\Omega_{ij}\equiv\pi_{ij}+
\frac{1}{2m}\epsilon_{ijk}A^k\approx 0
\end{eqnarray}
along with their associated Lagrange multipliers $\lambda^0$,
$\lambda^i$, $\Sigma^{0i}$, and $\Sigma^{ij}$.

There are four additional secondary constraints which are obtained
from the time stability conditions of the primary constraints
$\pi_0$ and $\pi_{0i}$ as
\begin{eqnarray}
\label{sec-con} \Lambda&\equiv&
A_0+\frac{1}{2m}\epsilon_{ijk}\partial^iB^{jk}\approx
0,\nonumber\\
\Lambda_i&\equiv& B_{0i}-\frac{1}{m}\epsilon_{ijk}\partial^j
A^k\approx 0.
\end{eqnarray}
As expected, the Lagrange multipliers $\Sigma^{ij}$, $\lambda^i$
corresponding to the constraints $\Omega_i$, $\Omega_{ij}$ are
fixed under the time stability condition, and the other ones
$\lambda^0$, $\Sigma^{0i}$ are determined by the consistency
requirement of the secondary constraints $\Lambda$, $\Lambda_i$,
respectively.

Then, the full set of these constraints makes the constraint
algebra second-class as
\begin{eqnarray}
\label{con-algebra}
\{\pi_0, \Lambda\}&=& -\delta(x-y),\nonumber\\
\{\pi_{0i}, \Lambda_j \}&=& -\delta_{ij}\delta(x-y),\nonumber\\
\{\Omega_i, \Omega_{jk} \}&=&
-\frac{1}{m}\epsilon_{ijk}\delta(x-y),
\end{eqnarray}
where we have redefined the secondary constraints as
$\Lambda+\partial^i\Omega_i \rightarrow \Lambda
=\partial^i\pi_i+A^0+\frac{1}{4m}\epsilon^{ijk}\partial_iB_{jk}\approx
0$, and $\Lambda_i+\partial^j\Omega_{ij} \rightarrow \Lambda_i
=\partial^j\pi_{ij}+B_{0i}-\frac{1}{2m}\epsilon_{ijk}
\partial^jA^k\approx 0$. These new definitions make the constraint
algebra have no derivative terms. As a result, we have easily obtained
the following Dirac Brackets
\begin{eqnarray} \label{dirac-bra} \{A^0(x), A^i(y)\}_{D} &=&
\partial^i_x \delta(x-y), \nonumber \\
\{A^i(x), B^{jk}(y)\}_{D} &=& -m\epsilon^{ijk}\delta(x-y),
\nonumber\\
\{B^{0i}(x), B^{jk}(y)\}_{D} &=& (\delta^{ij}\partial^k_x -
\delta^{ik} \partial^j_x) \delta(x-y),\nonumber\\
\{A^i(x), \pi_j(y)\}_{D} &=& \frac{1}{2}\delta^i_j
\delta(x-y),\nonumber\\
\{A^0(x), \pi_{ij}(y)\}_{D} &=&
-\frac{1}{2m}\epsilon_{ijk}\partial^k_x \delta(x-y),
\nonumber\\
\{\pi_i(x), B_{0j}(y)\}_{D} &=&
\frac{1}{2m}\epsilon_{ijk}\partial^k_x
 \delta(x-y),\nonumber\\
 \{\pi_i(x), \pi_{jk}(y)\}_{D} &=& \frac{1}{4m}\epsilon_{ijk} \delta(x-y),
\nonumber\\
\{B^{ij}(x), \pi_{kl}(y)\}_{D} &=&
\frac{1}{2}(\delta^i_k\delta^j_l-\delta^j_k\delta^i_l)
 \delta(x-y)
\end{eqnarray}
for the massive vector-tensor theory with topological coupling.

Now, let us briefly recapitulate the BFT Hamiltonian embedding for
this theory. In order for that purpose, we have introduced
auxiliary fields having involutive relations in which not only
modified new constraints in the enlarged space are strongly
vanishing with each other, but also they have the vanishing
Poisson brackets, not the Dirac brackets, with physical quantities
such as Hamiltonian and fields themselves. Through this BFT
prescription, after introducing auxiliary fields paired as
$(\theta, \pi_\theta)$, $(Q^i, P_i)$, and $(\Phi^i, \Phi^{jk})$,
we have obtained the strongly involutive primary
\begin{eqnarray}
\label{involutive-constraints} \widetilde{\pi}_0 &=& \pi_0
+\theta,~~~
\widetilde{\pi}_{0i} = \pi_{0i} + Q_i, \nonumber\\
\widetilde{\Omega}_i &=& \Omega_i + \Phi_i, ~~~
\widetilde{\Omega}_{ij} = \Omega_{ij}+ \frac{1}{m}\Phi_{ij},
\end{eqnarray}
and secondary constraints
\begin{eqnarray}
\widetilde{\Lambda} &=& \Lambda + \pi_\theta =
A_0+\frac{1}{2m}\epsilon_{ijk}\partial^iB^{jk}+ \pi_\theta,
\nonumber\\
\widetilde{\Lambda}_i &=& \Lambda_i + P_i =
B_{0i}-\frac{1}{m}\epsilon_{ijk}\partial^j A^k+ P_i.
\end{eqnarray}
Moreover, the strongly involutive physical fields are also
obtained as
\begin{eqnarray}
\label{physfield} \widetilde{A}^0 &=& A^0 + \pi_\theta,~~~~
\widetilde{A}^i = A^i + \partial^i\theta -
\frac{1}{2}\epsilon^{ijk}\Phi_{jk}, \nonumber \\
\widetilde{\pi}_0 &=& \pi_0 + \theta,~~~~~~ \widetilde{\pi}_i =
\pi_i - \frac{1}{2m}\epsilon_{ijk}\partial^j
Q^k + \frac{1}{2}\Phi_i, \nonumber \\
\widetilde{B}^{0i} &=& B^{0i}+P^i,~~~ \widetilde{B}^{ij} =
B^{ij}-(\partial^iQ^j-\partial^jQ^i)+
m\epsilon^{ijk}\Phi_k, \nonumber\\
\widetilde{\pi}_{0i} &=& \pi_{0i}+Q_i,~~~~ \widetilde{\pi}_{ij} =
\pi_{ij}-\frac{1}{2m}\epsilon_{ijk}
\partial^k \theta + \frac{1}{2m}\Phi_{ij}.
\end{eqnarray}
Note that all these physical fields are terminated in the first
order of the auxiliary fields, and the Poisson brackets of these
fields in the extended phase space are exactly same as the Dirac
brackets (\ref{dirac-bra}) in the original phase space.

On the other hand, by using the strongly involutive physical fields
(\ref{physfield}), we have also obtained the extended canonical
Hamiltonian as
\begin{eqnarray}
\label{ext-cH} \widetilde{\cal H}_c &=&
\frac{1}{4}(B_{ij}-Q_{ij})^2+\frac{1}{2}m\epsilon_{ijk}(B^{ij}-Q^{ij})\Phi^k
-\frac{1}{2}m^2\Phi_i\Phi^i \nonumber\\
&-&\frac{1}{2}(A_i+\partial_i\theta)^2+\frac{1}{2}\epsilon_{ijk}
(A^i+\partial^i\theta)\Phi^{jk}+\frac{1}{4}\Phi_{ij}\Phi^{ij}
\nonumber\\
&+&\frac{1}{2}(B_{0i}+P_i)^2-\frac{1}{m}\epsilon_{ijk}(B^{0i}+P^i)
\partial^jA^k-\frac{1}{m}(B_{0i}+P_i)\partial_j\Phi^{ij}
\nonumber\\
&-&\frac{1}{2}(A_0+\pi_\theta)^2-\frac{1}{2m}(A_0+\pi_\theta)\epsilon_{ijk}
\partial^iB^{jk}+(A_0+\pi_\theta)\partial_i\Phi^i, \nonumber\\
&+& \pi_\theta \widetilde{\Lambda}+P_i\widetilde{\Lambda}_i,
\end{eqnarray}
where the last two terms are added to generate the Gauss'
constraints corresponding to the constraints, $\widetilde{\pi}_0$,
$\widetilde{\pi}_{0i}$, consistently. Then, the generating
functional for the extended first-class systems is given by
\begin{eqnarray}
\label{path-int} {\cal Z}&=&\int {\cal D}A^\mu {\cal D}\pi_\mu
{\cal D}B^{\mu\nu} {\cal D}\pi_{\mu\nu} {\cal D}\theta {\cal
D}\pi_\theta {\cal D}Q^i {\cal D} P_i {\cal D}\Phi^i {\cal
D}\Phi^{ij}\nonumber\\
&\times&\delta(\widetilde{\varphi}_\alpha)\delta(\Gamma_\beta)~
{\rm det}\mid\{\widetilde{\varphi}_\alpha, \Gamma_\beta \}\mid
e^{iS},
\end{eqnarray}
where
\begin{eqnarray}
S&=&\int d^4x ~\left[\pi_\mu\dot{A}^\mu+\pi_{0i}\dot{B}^{0i}
+\frac{1}{2}\pi_{ij}\dot{B}^{ij}+\pi_\theta\dot{\theta}+P_i\dot{Q}^i
+\frac{1}{2}\epsilon_{ijk}\Phi^i\dot{\Phi}^{jk}-\widetilde{\cal
H}_c \right],\nonumber\\
\end{eqnarray}
and $\Gamma_\alpha$ are appropriate gauge fixing functions which
have nonvanishing Poisson brackets with the modified first-class
constraints $\tilde{\varphi}_\alpha=(\widetilde{\pi}_0,
\widetilde{\pi}_{0i},\widetilde{\Lambda},\widetilde{\Lambda}_i,
\widetilde{\Omega}_i,\widetilde{\Omega}_{ij})$.

Now, by performing the momenta integrations in the generating
functional as usual \cite{hkp}, we have finally obtained the gauge
invariant Lagrangian corresponding to ${\cal \widetilde{H}}_c$ as
\begin{equation}\label{generating-ftn}
{\cal Z}=\int {\cal D}A^\mu {\cal D}B^{\mu\nu} {\cal D}\theta
{\cal D}Q^\mu {\cal D}\Phi^i {\cal D}\Phi^{ij}
\delta(Q^0)\delta(\Gamma_\beta)
det\mid\{\widetilde{\varphi}_\alpha, \Gamma_\beta \}\mid e^{iS_T},
\end{equation}
where
\begin{eqnarray}\label{tot-act}
S_T & = & \int d^4x~\left( {\cal L}_{GE} + {\cal
L}_{NWZ}\right), \\
\label{stuckelberg} {\cal L}_{GE} &=& -\frac{1}{4}
(B_{\mu\nu}-Q_{\mu\nu})^2 +\frac{1}{2}(A_\mu+\partial_\mu\theta)^2
+ \frac{1}{2m}\epsilon_{\mu\nu\rho\sigma} B^{\mu\nu}\partial^\rho
A^\sigma, \nonumber \\
\label{nwz}{\cal L}_{NWZ} &=&
\frac{1}{2}\epsilon_{ijk}\Phi^i\dot{\Phi}^{jk}+\left[F_{i0}-
\frac{m}{2}\epsilon_{ijk} (B^{jk}-Q^{jk})+
\frac{m^2}{2}\Phi_i\right]\Phi^i \nonumber\\
&-&\left[\frac{1}{2}\epsilon_{ijk}(A^i+\partial^i\theta)+\frac{1}{m}\partial_k
B_{0j}+\frac{1}{2m}\partial_0
B_{jk}+\frac{1}{4}\Phi_{ij}\right]\Phi^{jk},
\end{eqnarray}
where ${\cal L}_{GE}$ is the gauge embedded Lagrangian and ${\cal
L}_{NWZ}$ is a new type of the WZ Lagrangian. Note that we have
introduced the delta functional of a variable $Q^0$ in the measure
which serves the final gauge embedded Lagrangian ${\cal L}_{GE}$
manifestly covariant.

On the other hand, the infinitesimal gauge transformations for the
fields are given as
\begin{eqnarray}
\label{fin-ext-gt} \delta A^\mu &=&
-\partial^\mu\epsilon_A+\delta^\mu_j \epsilon^j_A,~~ \delta
B^{\mu\nu}=\partial^\mu\epsilon^\nu_B -
\partial^\nu\epsilon^\mu_B+(\epsilon^{kl}_B-\epsilon^{lk}_B)
\delta^\mu_k\delta^\nu_l, \nonumber\\
\delta\theta &=& \epsilon_A, ~~~~~~~~~~~~~~~~~\delta Q^\mu =
\epsilon^\mu_B,
\nonumber\\
\delta\Phi^i&=&\frac{1}{m}\epsilon^{ijk}\epsilon^B_{jk},~~~~~~~~~
\delta\Phi^{ij}=-\epsilon^{ijk}\epsilon^A_k
\end{eqnarray}
from the generator of the gauge transformation\footnote{See
Ref.\cite{KKP} for details.}, the first-class constraints
(\ref{involutive-constraints}). The transformations related to the
gauge symmetry are exactly same as the ones (\ref{zeromode2})
obtained from the symplectic Lagrangian scheme. On the other hand,
the other parameters $\epsilon^i_A$, $\epsilon^{ij}_B$ in Eq.
(\ref{fin-ext-gt}) are not related with the gauge symmetry. These
parameters are associated with the symplectic structure of the
topological coupling, which are absent in the symplectic
Lagrangian scheme. Since in the latter scheme the topological
coupling term is already of first ordered, we have no need to
introduce the auxiliary fields (momenta) which have become the
gauge generators in the former scheme. As a result, the new type
of the WZ Lagrangian, which may lack manifest covariance, is
indeed invariant under the extended transformations including
these parameters. Moreover, by fixing unitary gauge conditions
such as $\Phi^i=\Phi^{ij}=0$, the new type of the WZ Lagrangian
identically vanishes, while the strongly involutive constraints
$\tilde\Lambda$, $\tilde{\Lambda}^i$ in Eq.
(\ref{involutive-constraints}) become exactly same with the true
ones $\phi_{\epsilon_1}$, $\phi_{\vec\epsilon_2}$ in Eq.
(\ref{constsym}) generated from the symplectic embedding
procedure, respectively.

It is also important to note that in the generating functional
(\ref{generating-ftn}) there exists the delta functional of the
variable $Q^0$ which transforms as $\delta Q^0=\epsilon^0_B$. We
have introduced this new field to make the final Lagrangian
manifestly covariant. Even without this $Q^0$ field, we can show
that the resulting Lagrangian successfully reproduces all the BFT
embedded constraint structure. However, in that case, it will lose
manifest covariance. Indeed, we have introduced the new variable
$Q^0$ in order for keeping the manifest covariance, while giving
up the irreducible property between the constraints \cite{KKP}.

\section{Conclusion}

In this paper we have quantized a massive vector-tensor theory
with topological coupling which is a first order master action of
the Proca and the Kalb-Ramond models. We have adopted the
symplectic scheme since it provides relatively simple way of
getting the Dirac brackets rather than that of Dirac for the
Lagrangian having seemingly complicated antisymmetric tensor
fields. In particular, we have shown that our model has the only
four constraints in the symplectic scheme in contrast to the Dirac
(or, BFT) formalism having the fourteen constraints. Moreover, we
have demonstrated how the BFT embedding scheme of second-class
system into first-class one can be realized in the framework of
the symplectic approach to constrained system. Rather than
proceeding iteratively as in the BFT embedding approach, we have
greatly simplified the calculation by making use of manifest
Lorentz invariance in our {\it Ansatz} for the WZ term.

Furthermore, we have explicitly shown that the reducibility
between the constraints in the resulting gauge invariant
Lagrangian comes from the absence of the $Q^0$ term, which
naturally arises in analyzing the model's symplectic structure.
Requiring the gauge invariance of the total Lagrangian, we have
successfully recovered it and thus have obtained the full gauge
transformations consistently, while showing that all the zero
modes solutions are not linearly independent, {\it i.e.,} there
exists the first-stage reducible trivial zero mode.

Similar to the symplectic Lagrangian embedding, we have shown in
the BFT Hamiltonian embedding that the reducibility between the
constraints is also related to the $Q^0$ component, and have
successfully reconstructed the Lorentz covariant gauge embedded
Lagrangian by using the delta functional of the $Q^0$ field in the
path integral measure. However, in this scheme which carries out
the rigorous converting procedure from the second-class
constraints to the effective first-class ones, due to the
appearance of the constraints $\Omega_i$, $\Omega_{ij}$ in Eq.
(\ref{involutive-constraints}) originated from the symplectic
structure of the theory, we can not keep the Lorentz covariance as
seen in the new type of the WZ Lagrangian. Nevertheless, we have
shown that the total action including the NWZ action is invariant
under the extended gauge transformations. Compared with this,
since in the symplectic Lagrangian embedding we do not need to
classify the constraints as the second-, or the first-, we could
freely require the Lorentz covariance without any inconsistency.
In short, gauge embedded extension in the symplectic scheme is
only related with the Lorentz covariant action while the usual
symplectic scheme concerns only with true constraints.

Finally, the method followed by this work is worthwhile in itself
because there is no known systematic method yet except the
simplest one-form cases \cite{kpkk,sym-emb,hkp}. In this respect,
we have newly generalized the symplectic Lagrangian embedding
procedure to include highly non-trivial tensor fields which
exhibit, for example, the existence of the first-stage reducible
trivial zero mode among the others related to the reducibility of
theory.

\vskip 1.0cm

The work of YWK was supported by the Korea Research Foundation,
Grant No. KRF-2002-075-C00007. The work of CYL was supported by
KOSEF, Grant No. R01-2000-000-00022-0.

\end{document}